\pdfoutput=1
\documentclass{article}

\usepackage[symbol]{footmisc}
\usepackage{amssymb,amsfonts,amsmath,stmaryrd}
\usepackage{color}

\def\be{\begin{eqnarray}}
\def\ee{\end{eqnarray}}
\def\nn{\nonumber}

\def\tr{{\rm tr}\,}
\def\Tr{{\rm Tr}\,}

\newcommand{\beq}{\begin{equation}}
\newcommand{\eeq}{\end{equation}}
\newcommand{\beqa}{\begin{eqnarray}}
\newcommand{\eeqa}{\end{eqnarray}}

\definecolor{red}{rgb}{1,0,0}
\definecolor{orange}{rgb}{1,0.5,0}
\definecolor{violet}{rgb}{0.7,0,1}

\hoffset= - 1.0in         

\begin{document}
\begin{center}
\begin{small}
\hfill FIAN/TD-02/21\\
\hfill IITP/TH-02/21\\
\hfill ITEP/TH-02/21\\
\hfill MIPT/TH-02/21\\
\end{small}
\end{center}

\vspace{.5cm}

\begin{center}
\begin{Large}\fontfamily{cmss}
\fontsize{17pt}{27pt}
\selectfont
	\textbf{ Superintegrability and Kontsevich-Hermitian  relation}
	\end{Large}
	
\bigskip \bigskip
\begin{large}A. Mironov$^{a,b,c}$\footnote{mironov@lpi.ru; mironov@itep.ru},
A. Morozov$^{d,b,c}$\footnote{morozov@itep.ru}
 \end{large}
\\
\bigskip

\begin{small}
$^a$ {\it Lebedev Physics Institute, Moscow 119991, Russia}\\
$^b$ {\it ITEP, Moscow 117218, Russia}\\
$^c$ {\it Institute for Information Transmission Problems, Moscow 127994, Russia}\\
$^d$ {\it MIPT, Dolgoprudny, 141701, Russia}\\
\end{small}
 \end{center}
\medskip

\begin{abstract}
We analyze the well-known equivalence between
the quadratic Kontsevich-Penner and Hermitian matrix models
from the point of view of superintegrability relations,
i.e. explicit formulas for character averages.
This is not that trivial on the Kontsevich side,
and seems important for further studies of various deformations
of Kontsevich models.
In particular, the Brezin-Hikami extension of the above equivalence
becomes straightforward.
\end{abstract}

\section{Introduction}

According to \cite{Che,UFN3,AMM},
the Hermitian matrix model is equivalent to the quadratic Kontsevich-Penner model
in the following sense:
\be
\int_{N\times N} \exp\left(\sum_k \frac{p_k}{k}\Tr M^k\right)\ e^{-{1\over 2}\Tr M^2} dM
\sim \int_{n\times n}  {\det}^{- N}\!(1-L^{-1}X) \ e^{-{1\over 2}\tr X^2} dX
\label{hekoid}
\ee
with
\be
p_k = \tr L^{-k}
\label{pkL}
\ee
and the Gaussian integrals normalized to unity,
$\int e^{-{1\over 2}\Tr M^2} dM = \int e^{-{1\over 2}\tr X^2} dX = 1$.
Our goal in this note is to make  this identity compatible with
the superintegrability property \cite{IMMsi},
\be
\int_{N\times N}  \chi_R[M] \ e^{-{1\over 2}\Tr M^2} dM
= \frac{\chi_R\{\delta_{k,2}\}}{\chi_R\{\delta_{k,1}\}}\cdot \chi_R\{N\}
\label{si}
\ee
which provides explicit expressions
for averages of the Schur functions $\chi_R[M]$ on the both sides. Here the Schur function \cite{Mac} is a symmetric function of eigenvalues of the matrix $M$, and it is understood as a function of power sums of the eigenvalues, $\chi_R[M] := \chi_R\{\Tr M^k\}$.
In the Kontsevich case, these averages were not yet discussed in the literature
but are somehow similar to the ones appearing in the cubic Kontsevich \cite{MM3ko}  (see also \cite{dFIZ})
and generalized Kontsevich models \cite{MMgkm}, and in the Brezin-Gross-Witten model \cite{Al}. For more examples, see
\cite{IMMsi,MMten,PSh}.

\section{The basic formula}

Our consideration is based on a simple generalization of (\ref{si}) to the Gaussian Hermitian model in the external field with partition function
\be\label{Z}
Z=\int_{N\times N} e^{-{1\over 2}\Tr MLML+\sum_k g_k\Tr M^k} dM
\ee
where $M$ is the Hermitian $N\times N$ matrix, $dM$ is the Haar measure, $g_k$ are parameters, and the integral is understood as a formal power series in these parameters $g_k$. The correlators are defined
\be
<F(M)>_L:=\int_{N\times N} F(M)e^{-{1\over 2}\Tr MLML} dM
\ee
and we normalize the measure in such a way that $<1>\ =1$.

The average of the Schur functions is easily performed in this model with the help of the Wick theorem following the line of \cite{MMten,MM3ko}, and it gives
\be
\boxed{
\Big<\chi_R[M] \Big>_L
= c_R\cdot\chi_R[ L^{-1}]
}
\label{gav}
\ee
with
\be
c_R = \frac{\chi_R\{\delta_{k,2}\}}{\chi_R\{\delta_{k,1}\}}
\label{cR}
\ee

\bigskip

This relation (\ref{gav}) is a direct analogue of the result of \cite{MM3ko} for the rectangular complex matrix model,
which further develops its analogy with the Kontsevich family.

\section{The main relation}

There are many proofs of formula (\ref{hekoid}), see, for instance, \cite{AMM}. Here we are going to derive (\ref{hekoid}) using the basic formula (\ref{gav}). To this end, we make a change of variables $M\to L^{-1}M$ in the integral (\ref{Z}) so that the averages in (\ref{gav}) are evaluated with the Gaussian weight $e^{-{1\over 2}\Tr M^2}$, but the Schur functions become depending on $L^{-1}M$:
\be
\boxed{
\Big<\chi_R[L^{-1}M] \Big>
= c_R\cdot\chi_R[ L^{-1}]
}
\label{gav1}
\ee
where $<\ldots>$ is understood as the average with the Gaussian measure (\ref{si}), i.e. without the external matrix.

At the l.h.s. of (\ref{hekoid}), it is sufficient to apply the Cauchy formula \cite{Mcau}
\be
\exp\left(\sum_k \frac{p_kp'_k}{k}\right) = \sum_R \chi_R\{p\}\chi_R\{p'\}
\ee
where the sum goes over all Young diagrams $R$, in order to get
\be
\int_{N\times N}  \exp\left(\sum_k \frac{p_k}{k}\Tr M^k\right)\ e^{-{1\over 2}\Tr M^2} dM
= \sum_R \chi_R\{p\} \int_{N\times N}  \chi_R[M] \ e^{-{1\over 2}\Tr M^2} dM
\ee
Applying the result of \cite{IMMsi} (formula (\ref{si}) above), we get for the l.h.s. of (\ref{hekoid})
\be\label{si2}
\int_{N\times N}  \exp\left(\sum_k \frac{p_k}{k}\Tr M^k\right)\ e^{-{1\over 2}\Tr M^2} dM
= \sum_R \frac{\chi_R\{p\}\chi_R\{N\}\chi_{R}\{\delta_{k,2}\} }{\chi_R\{\delta_{k,1}\}}
\ee

Now consider the r.h.s. of (\ref{hekoid}). In \cite{Che,UFN3}, the determinant in this formula entered in a positive degree (see, e.g., \cite[Eqs.(I.1),(I.17)]{AMM}), which gave rise to a simple definition of the integral, since it was just the Gaussian average of a polynomial. It was for the price of various imaginary units and of the minus sign in (\ref{pkL}). Here we consider the integral at the r.h.s. of (\ref{hekoid}) as a power series at large $L$:
\be
\!\!
\int_{n\times n}\!\!\! {\det}^{- N}\!(1-L^{-1}X) \ e^{-{1\over 2}\tr X^2} dX
= \int_{n\times n}\!\!\!\exp\left( N \sum_k \frac{1}{k}\tr (L^{-1}X)^k\right)   \ e^{-{1\over 2}\tr X^2} dX
\nn
\ee
Applying once again the Cauchy formula, we obtain
\be
\!\!
\int_{n\times n}\!\!\! {\det}^{- N}\!(1-L^{-1}X) \ e^{-{1\over 2}\tr X^2} dX
=\sum_R \chi_R\{N\}\,\Big<\chi_R[L^{-1}X] \Big>
\nn
\ee
Using now (\ref{gav1}), we finally come to
\be
\boxed{
\int_{n\times n}  {\det}^{- N}\!(1-L^{-1}X) \ e^{-{1\over 2}\tr X^2} dX =
\sum_R \frac{\chi_R[L^{-1}]\chi_R\{N\}\chi_{R}\{\delta_{k,2}\} }{\chi_R\{\delta_{k,1}\}}\stackrel{(\ref{si2})}{=}
\int_{N\times N}  \exp\left(\sum_k \frac{p_k}{k}\Tr M^k\right)\ e^{-{1\over 2}\Tr M^2} dM
}\nn\\
\label{toprove}
\ee

\bigskip

\paragraph{An example.}
As a small illustration of how this works, in the first approximation to (\ref{toprove}), we have
\be
\int \frac{ N\tr (L^{-1}XL^{-1}X) +  N^2(\tr L^{-1}X)^2}{2}  \ e^{-{1\over 2}\tr X^2} dX
= \frac{N^2p_2 + Np_1^2}{2 }
\nn
\ee
where we used the relation
\be
\left<X_{ij} X_{kl}\right>\ = \frac{\delta_{jk}\delta_{il}}{n^2}\left<\tr X^2\right> \ = \delta_{jk}\delta_{il}
\nn
\ee
for the Gaussian correlators.
At the same time,
\be
\sum_{R:\ |R|=2} \frac{\chi_R\{\tr L^{-k}\}\chi_R\{N\}\chi_{R}\{\delta_{k,2}\} }{\chi_R\{\delta_{k,1}\}}
= \chi_{[2]}\{\tr L^{-k}\}\chi_{[2]}\{N\} - \chi_{[1,1]}\{\tr L^{-k}\}\chi_{[1,1]}\{N\}
= \nn \\
= \frac{N(N+1)(p_2+p_1^2) - N(N-1)(-p_2+p_1^2)}{4} = \frac{N^2 p_2 + Np_1^2}{2}
\nn
\ee

\section{Brezin-Hikami identity }

Relation (\ref{gav}) allows one to write a more symmetric version of (\ref{toprove}),
which appears to be the Brezin-Hikami generalization \cite{BH1,BH}
of the Chekhov-Makeenko relation (\ref{hekoid}).
To emphasize the symmetry, we first write it in a slightly different notation:
\be
\boxed{
\int_{N_1\times N_1} {\rm Det}^{-1}\Big(1- L_1^{-1}M_1\otimes L_2^{-1}\Big)\cdot e^{-{1\over 2}\Tr_1 M_1^2}dM_1
= \int_{N_2\times N_2} {\rm Det}^{-1}\Big(1- L_1^{-1} \otimes L_2^{-1}M_2\Big)\cdot e^{-{1\over 2}\Tr_2 M_2^2}dM_2
}
\label{symid}
\ee
In this identity, there are determinants of the tensor product, and we write $1$ instead of $Id\otimes Id$
in order to simplify the notation.
By the Cauchy formula, the inverse determinant at the l.h.s. is equal to
\be
{\rm Det}^{-1}\Big(1- L_1^{-1}M_1\otimes L_2^{-1}\Big)
= \exp\left(\sum_k \frac{1}{k} \Tr\!_1(L_1^{-1}M_1)^k \,\Tr\!_2 L_2^{-k}\right)
= \sum_{R} \chi_R\Big\{\Tr\!_1(L_1^{-1}M_1)^k\Big\} \chi_R\Big\{\Tr\!_2 L_2^{-k}\Big\}
\ee
and the l.h.s. itself becomes
\be
\int_{N_1\times N_1} {\rm Det}^{-1}\Big(1- L_1^{-1}M_1\otimes L_2^{-1}\Big)\, e^{-{1\over 2}\Tr\!_1 M_1^2}dM_1
= \sum_{R}  \Big<\chi_R\left\{\Tr\!_1(L_1^{-1}M_1)^k\right\}\Big>_{N_1\times N_1}
\cdot \chi_R\left\{\Tr\!_2 L_2^{-k}\right\}
= \nn
\ee
\vspace{-0.5cm}
\be
\ \stackrel{(\ref{gav})}{=}\
\sum_R c_R\cdot \chi_R\left\{\Tr\!_1 L_1^{-k}\right\}\chi_R\left\{\Tr\!_2 L_2^{-k}\right\}
\ee
Clearly, the r.h.s. of (\ref{gav}) is just the same because of the symmetry $1\leftrightarrow 2$, and this proves the relation (\ref{symid}).

\bigskip

\paragraph{An example.}
It is again instructive to look at the first  approximation to (\ref{symid}).
At the l.h.s., we have a clearly symmetric expression:
\be
\int_{N_1\times N_1} \frac{ \Tr\!_1 (L_1^{-1}M_1L_1^{-1}M_1)\cdot\Tr\!_2 L_2^{-2}
+  (\Tr\!_1 L^{-1}M_1)^2 \cdot (\Tr\!_2 L_2^{-1})^2}{2}  \ e^{-{1\over 2}\Tr\!_1 M_1^2} dX=\nn\\
= \frac{(\Tr\!_1 L_1^{-1})^2\cdot\Tr\!_2 L_2^{-2}+ \Tr\!_1 L_1^{-2}\cdot(\Tr\!_2 L_2^{-1})^2}{2 }
\nn
\ee
where we used
\be
\left<M_{ij} M_{kl}\right>\ = \frac{\delta_{jk}\delta_{il}}{N^2}\left<\Tr M^2\right> \ = \delta_{jk}\delta_{il}
\nn
\ee
for the Gaussian correlators.

\section{Conclusion}

In this paper, we exploited a new relation (\ref{gav}) for the Gaussian averages,
and demonstrated that it stands behind the Chekhov-Makeenko identity (\ref{hekoid})
between the Hermitian and quadratic Kontsevich-Penner models, and behind its further Brezin-Hikami extension.

In addition to these ``practical" applications, one can also wonder about the theoretical
meaning of our results.
{\bf The superintegrability property (\ref{si}) is a direct corollary of (\ref{gav})} at $L=1$,
with the obvious substitution of the $n\times n$ matrix $X$ by the $N\times N$ matrix $M$.
This can make (\ref{gav}) a reasonable enhancement of (\ref{si}).
However, the inverse claim, i.e. whether (\ref{gav}) is directly implied by (\ref{si}), remains unclear.
A possible approach to this problem can require supplementing (\ref{si}) with some version of
the Wick theorem which can be used as a consistency relation for (\ref{si}) and a source of
some stronger statements like (\ref{gav}).
Alternatively, one can just postulate (\ref{gav}) {\it in addition} to (\ref{si}),
but this requires a study of possible mutual restrictions on the two definitions.
A separate interesting question is the relation between (\ref{si}), (\ref{gav})
and factorization property of single-trace Harer-Zagier functions \cite{HZ,HZ1}.
All these issues remain for the future work.

\section*{Acknowledgements}

This work was supported by the Russian Science Foundation (Grant No.20-12-00195).

\end{document}